\documentclass[11pt]{article}
\usepackage[fontsize=11.0pt]{scrextend}
\usepackage{float}
\restylefloat{table}
\usepackage{graphicx}
\usepackage{rotating}
\usepackage{color}
\usepackage[dvipsnames]{xcolor}
\usepackage{enumitem}
\usepackage{sidecap}
\usepackage[top=8mm, bottom=8mm, left=8mm, right=8mm]{geometry}
\usepackage{multicol}
\usepackage[dvipsnames]{xcolor}
\usepackage{wrapfig}
\usepackage{caption}
\usepackage{afterpage}
\usepackage[export]{adjustbox}
\usepackage[official]{eurosym}
\usepackage[symbol]{footmisc}
\usepackage[round]{natbib}   
\usepackage{amssymb}

\usepackage{fancybox}

\begin{document}

\title{White Paper Stellar Mass}

\begin{center}

A white paper for the ESO Expanding Horizons initiative

\vspace{4mm}

\huge {\textbf{Decomposing the growth mechanisms of galaxies over the last 10\,billion years}}\\

\vspace{-2mm}

\normalsize

\vspace{4mm}

\Large{Luke J. M. Davies\footnote[2]{Email: luke.j.davies@uwa.edu.au}$^1$,  Annagrazia Puglisi$^2$,  Marcella Longhetti$^3$, Mark Sargent$^4$, Simon P. Driver$^1$, Aaron S. G. Robotham$^1$, Sabine Bellstedt$^1$, Fabio Rosario Ditrani$^{3,5}$, Anna R. Gallazzi$^{6}$, Laura Scholz Díaz$^{6}$, Stefania Barsanti$^{7}$, Stefano Zibetti$^{6}$, Sabine Thater$^{8}$}\\

\end{center}

\vspace{3mm}

\normalsize
\begin{center}
\begin{minipage}[t]{0.9\textwidth}
\textit{$^{1}$ ICRAR, The University of Western Australia, 35 Stirling Highway, Crawley, WA 6009, Australia \\
$^{2}$ School of Physics and Astronomy, University of Southampton, Highfield, SO17 1BJ, Southampton, UK \\
$^{3}$ INAF-Osservatorio Astronomico di Brera, via Brera 28, I-20121 Milano, Italy\\
$^{4}$ EPFL Laboratory of Astrophysics (LASTRO), Observatoire de Sauverny, CH – 1290 Versoix, Switzerland \\
$^{5}$ Universit\`{a}degli Studi di Milano-Bicocca, Piazza della Scienza, I-20125 Milano, Italy\\
$^{6}$ NAF-Osservatorio Astrofisico di Arcetri, Largo Enrico Fermi 5, 50126, Firenze, Italy\\
$^{7}$ Sydney Institute for Astronomy, School of Physics, University of Sydney, NSW 2006, Australia \\
$^{9}$ Department of Astrophysics, University of Vienna, Türkenschanzstrasse 17, 1180 Vienna, Austria\\
}
\end{minipage}
\end{center}
\normalsize

\vspace{1mm}

\newpage

\begin{center}
\Ovalbox{
\begin{minipage}{7.5in}
\begin{center}
\begin{minipage}{7.3in}
\vspace{1mm}

Determining how galaxies accumulate stellar mass is paramount to understanding the Universe. Two primary mechanisms drive this process: star-formation (SF) \& mergers. Our understanding of star formation, and to some degree the processes that influence the baryon cycle (environment, gas supply, feedback, etc), are either relatively well constrained or will develop significantly over the coming decades via upcoming facilities ($i.e.$ through their imprint on galaxy  properties measured with deep multi-wavelength and spectroscopic data). However, the same can not be said for mergers.  \textbf{\textcolor{teal}{It is telling that we indirectly know hierarchical assembly through mergers is one of the most crucial processes that shape our Universe, but the robust observational measurement of mergers is almost non-existent outside of the local Universe - let alone how these mergers impact galaxy properties.}} This is not likely to significantly change in the coming decades as existing or approved facilities/surveys are inadequate in charactering mergers in the distant Universe. Motivated by this, we discuss an ambitious study to first explore mergers, and then the co-dependent astrophysical process that govern the accumulation of stellar mass over the last $\sim$10\,billion years, and highlight the essential need for a 10m+ class multi-object spectroscopic facility.  

\vspace{1mm}
\end{minipage}
\end{center}
\end{minipage}
}
\end{center}

\vspace{-2mm}

A ubiquitous feature of the Universe is that galaxies grow more massive over time. They typically start as small irregular-like systems and over the last $\sim$13 Billion years grow into the colossal structures we observe today, many orders of magnitude more massive ($e.g.$. \textcolor{blue}{Springer+ 2005}). To first order, this evolution is driven by two processes: star-formation and mergers, both of which lead to the formation/accumulation of stellar mass. However, the astrophysical processes underpinning these two processes are vast and complex. SFRs are typically determined by the availability of cold gas (\textcolor{blue}{Kennicutt 1998}), which in-turn underpins many factors which shape the evolution of galaxies: stellar mass (\textcolor{blue}{Popping+ 2014, Walter+ 2020}), metallicity (\textcolor{blue}{Casasola+ 2017}), morphology (\textcolor{blue}{Davies+ 2025b}), local (\textcolor{blue}{Davies+ 2025e}) and larger-scale environment (\textcolor{blue}{Davies+ 2019b, Davies+ 2025c}), galaxy-galaxy interactions (\textcolor{blue}{Davies+ 2015b, Pearson+ 2019}), the current strength of AGN- and/or Supernova-driven feedback (\textcolor{blue}{Davies+ 2025b}), etc. Likewise, merger rates of galaxies vary as a function of stellar mass (\textcolor{blue}{Robotham+ 2014}), epoch (\textcolor{blue}{Fuentealba-Fuentes+ 2025}), wet/dry fraction (\textcolor{blue}{Lin+ 2008}), and likely, large-scale-structure location (\textcolor{blue}{Neistein+ 2008}).

\vspace{1mm}

Over the last $\sim$20 yrs we have mapped galaxy growth via indirect tracers, e.g the changing distribution of stellar
mass  (Stellar Mass Function, \textcolor{blue}{Muzzin+ 2013, Thorne+ 2021}), gas masses (Gas mass function, \textcolor{blue}{Popping+ 2014}), sizes (\textcolor{blue}{Sargent+2007, Cook+ 2025}), morphologies (\textcolor{blue}{Driver+ 2022, Hashemizadeh+ 2021}), internal structure (\textcolor{blue}{Cook+ 2025}), SFRs (\textcolor{blue}{Thorne+ 2021}). \textbf{However, we currently know very little about the underlying astrophysics that lead to these relations and their evolution ($i.e.$ we know the \textit{correlations}, but not their \textit{causation}).} This is largely due to observational limitations in our ability to trace this astrophysics outside of all but the very local Universe - thus restricting any evolutionary baseline.     

\vspace{1mm}

Holistically, our science focus is an ambitious study to determine the impact of various astrophysical processes in shaping the accumulation of stellar mass in the Universe and map this to the evolution of the scaling relations noted above. This is summarised in Fig. 1. The left panels show processes that either increase growth in stellar mass (orange box) or impact the star-formation cycle slowing or stopping growth in stellar mass (purple box). If we are able to measure the impact of all of these processes on stellar mass growth and gas depletion at a given stellar mass and over and interval, $\Delta$t, we can map their relative contribution to the growth of galaxies (middle panel). In theory, if we are accounting for all processes the sum of these should match the evolution of distributions such as the Stellar Mass Function and the Gas Mass Function over the same time interval. Such a self-consistent and all-encompassing model of stellar mass growth, would require that we measure all of these properties consistently across a minimally-biased, volume-limited, sample of galaxies spanning a diverse range of epochs, environments and galaxy types.

\vspace{1mm}

Achieving this ambitious goal requires coordinated observations across multiple wavelength regimes (X-ray to radio) and techniques (imaging, spectroscopy, IFU/IFS),  and, critically, seamless combination of these data for core science programs. We have begun approaching this lofty goal first through the Galaxy and Mass Assembly survey (GAMA, \textcolor{blue}{Driver+ 2011}) and the Deep Extragalactic VIsible Legacy (DEVILS, \textcolor{blue}{Davies+ 2018, 2025e}), and are continuing with 4MOST’s Wide Area VISTA Extragalactic Survey (WAVES, \textcolor{blue}{Driver+ 2019}). 

\vspace{1mm}

However, even following WAVES in the 2030’s we will still only be able to explore the relative importance of star-
formation and mergers to shaping stellar mass growth at just $z$$<$0.45, and primarily for high-stellar-mass systems
undergoing major mergers (mass ratios $>$1:3) - let alone determine the underlying astrophysics. Beyond WAVES,
achieving transformative results in this area will require a new facility capable of delivering the complementary optical–NIR spectroscopic observations needed to identify and characterise mergers beyond the local Universe.

\vspace{-1mm}

\begin{figure*}[!h]
\begin{center}
\includegraphics[scale=0.4]{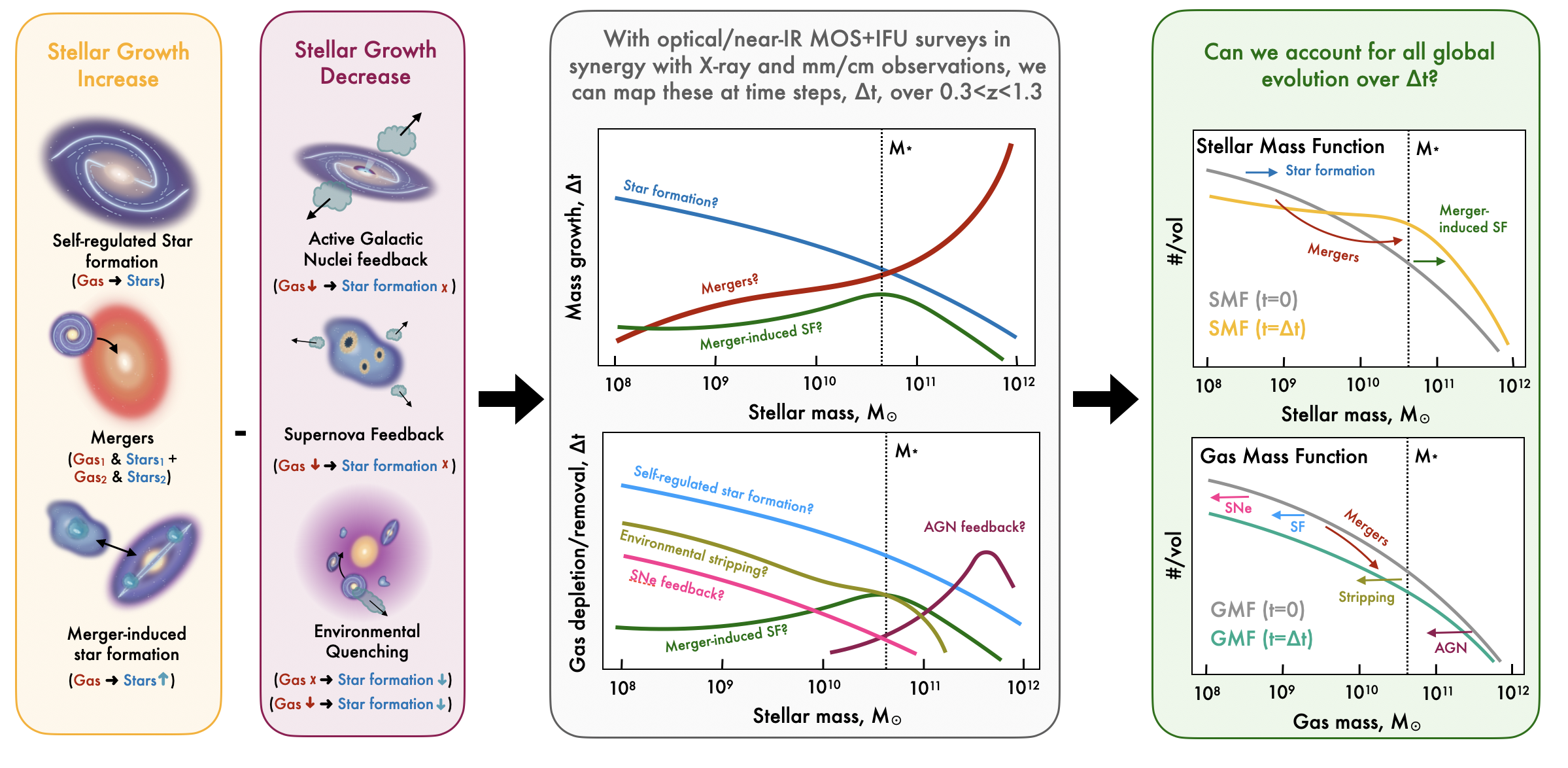}

\vspace{-5mm}

\caption{\textit{\small{Overview of the methodology used to identify the key drivers of stellar mass growth in galaxies. Left: Galaxies grow via star-formation and mergers, which rate is regulated by many astrophysical processes (e.g., AGN/SNe feedback, environmental quenching etc). Middle: By combining a deep redshift survey with synergistic studies from e.g. SKA we can measure the stellar mass growth and gas depletion rates over a time interval, $\Delta$t, decomposed into distinct astrophysical mechanisms. Right: If all processes are accounted for, the sum of these should account for the evolution in the SMF and GMF. Credit galaxy images: NASA}}}

\vspace{-11mm}

\label{fig:fig1}
\end{center}
\end{figure*}

\vspace{2mm}   

\textbf{WHAT IS MISSING?  } Over the coming decades, numerous surveys/facilities will drastically change our understanding of star formation and of the physical processes governing the baryon cycle over cosmological timescales, and provide these core aspects to Fig 1. Radio and millimetre facilities such as the SKA, ngVLA, and ALMA-WSU will deliver transformational insights into the cold gas content of galaxies, star formation, AGN activity, and feedback processes, while deep, high-resolution multi-wavelength surveys with Euclid and Roman will map galaxy morphologies and structural evolution across cosmic time. When coupled with deep, large-volume optical spectroscopic surveys (providing redshifts, AGN diagnostics, environmental metrics, and kinematics; see below), our picture of the accumulation of stellar mass via star formation, and the physical processes regulating it, will be fundamentally transformed.

\vspace{1mm}       

However, this only covers one of the processes that shape the distribution stellar mass we see today. \textbf{If we envisage how our understanding of mergers will progress in the coming decades, the future is more bleak.} 

\vspace{1mm}     

Identifying merger systems outside of the local Universe and/or to low stellar mass (and high merger ratios) is fraught with difficulty. Broadly speaking, we can only robustly understand how mergers shape the galaxy population if we identify \textit{all} mergers over a given volume to stringently-defined stellar mass and merger ratio limits (see \textcolor{blue}{Fuentealba-Fuentes+ 2025}). Without this, incompleteness and biases in merger samples limit their usefulness.  

\vspace{1mm}    

There are multiple methods for identifying mergers of galaxies: via visually-disturbed morphologies ($e.g.$ \textcolor{blue}{Conselice+ 2009} - requiring multi-wavelength high resolution imaging), post-starburst systems ($e.g.$ \textcolor{blue}{Wilkinson+ 2022} - requiring high SNR spectra), non-uniform internal dynamics (\textcolor{blue}{Loubser+ 2022} - requiring deep IFU/IFS studies), and the spectroscopic identification of close pairs ($e.g.$ \textcolor{blue}{Robotham+ 2014, Fuentealba-Fuentes+ 2025} - requiring highly complete spectroscopic samples), each with their own successes, failures and biases. If we wish to fully understand mergers and their impact on galaxies, we must explore all of these for same volumes. However, we argue here that the most fundamental method to exploring the redistribution of stellar mass, and the most complete approach, is that using pre-merger close pairs, where we can separate the properties (stellar mass, gas mass, SFR, etc) of the pre-merger products. These types of studies have proved highly successful in the local Universe ($i.e.$ using SDSS and GAMA). However, to date, observational constraints have severely impacted our ability to identify these close pairs outside of the local Universe due to the requirement of highly complete spectroscopic samples (Fig 2 - left \& middle). To highlight this, WAVES-deep will observe galaxies to 0.33\,M* out to $z$=0.8, but due to, even minimal, spectroscopic incompleteness, we predict that we will only be able to measure M* major merger rates to $z$$\sim$0.45. To go beyond this, we must obtain highly complete spectroscopic samples to much lower stellar masses than those obtained by WAVES. 

\vspace{1mm}

Importantly, mergers not only redistribute stellar mass but also significantly impact galaxy properties - shaping the baryon cycle itself (\textcolor{blue}{Darg+ 2010, Sargent+ 2025}). They can induce SF (\textcolor{blue}{Davies+ 2015b, Scudder+ 2012}), lead to quenching (\textcolor{blue}{Davies+ 2016a}), strip gas (\textcolor{blue}{Stevens+ 2021}), trigger AGN feedback-driven quenching (\textcolor{blue}{Ellison+ 2019}), shape the internal morphology, structure and dynamics of a galaxy (\textcolor{blue}{Martin+ 2018}), and determine a galaxy's overall long-term evolutionary path ($i.e.$ wet vs dry mergers, major vs minor merger rates, \textcolor{blue}{Lin+ 2008}). To date, we know little about these precesses outside of the local Universe. Only if we build robust samples of mergers, we can then begin to explore how mergers impact the underlying astrophysics within galaxies. Such studies would require the combination of high-resolution imaging, radio continuum imaging/spectra, deep and large volume IFS studies and large area deep and high completeness spectroscopic samples - with the latter two currently non-existent in upcoming approved facilities.

\begin{figure*}
\begin{center}
\includegraphics[scale=0.53]{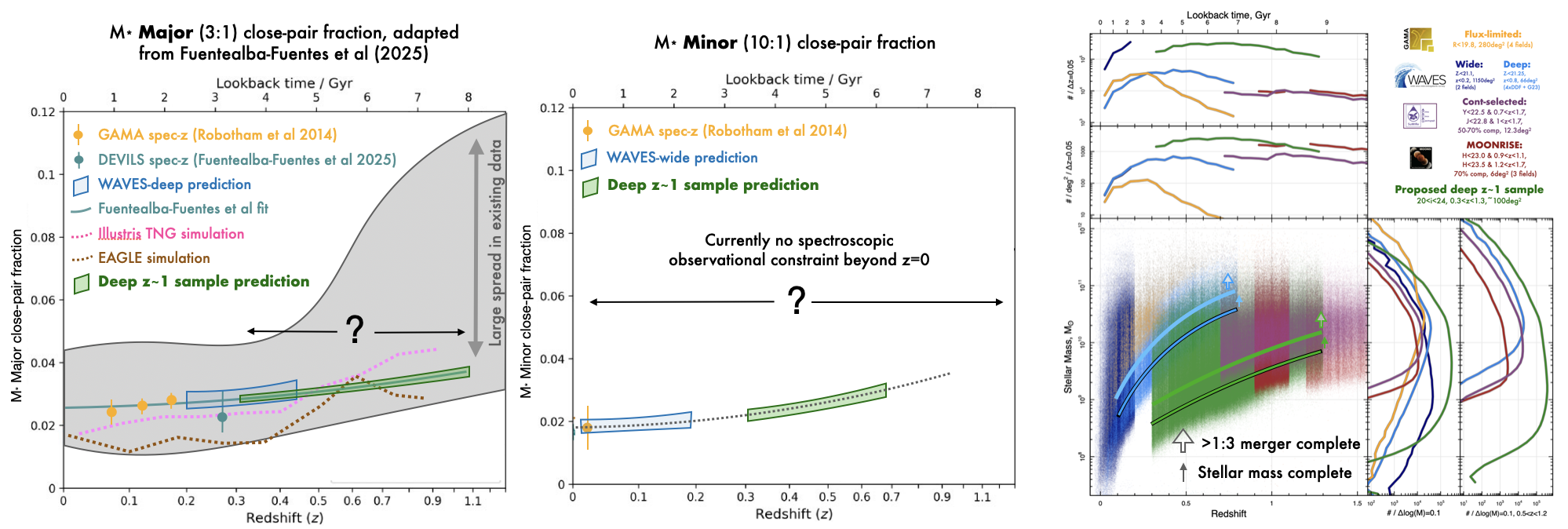}

\vspace{-3mm}

\caption{\textit{\small{Left: Robust measurements of even the major-merger M* close-pair merger fraction are almost non-existent outside of the local Universe. Middle: We can only extend this to minor mergers at z$\sim$0.  Right: proposed sample selection which would revolutionise our understanding of mergers (and many other science areas).}}  }

\vspace{-11mm}

\label{fig:fig2}
\end{center}
\end{figure*}

\vspace{2mm}

\textbf{SAMPLE AND STRATEGY  }Identifying mergers from close pairs requires highly-complete spectroscopic samples, coupled with extensive multi-wavelength data. This lends itself to using a large-area, deep galaxy redshift survey aligned with deep observations from other surveys/facilities ($i.e.$ WAVES, Roman, SKA). We also note that complementary large-area blind IFS data would allow such studies to be extended to lower stellar-mass and minor-merger regimes (mass ratios $<$1:3), and provide complementary dynamical identification of post-merger systems. Considering other upcoming surveys (Fig 2. right), WAVES-deep will be able to robustly probe major merger rates to M*  out to z$\sim$0.45, while WAVES-wide will be able to extend down to minor mergers, but only in the very local Universe. MOONRISE (\textcolor{blue}{Maiolino+ 2020}) and the PFS continuum-selected survey (\textcolor{blue}{Greene+ 2022}) will probe faint galaxy populations at z$>$0.7, but will likely sparse-sample ($\sim$50-70\%), rendering them inadequate to measure galaxy merger rates. Likewise, studies such as those with the ELT, will probe too small a volume to be representative. Finally we note, that spectroscopic redshifts are essential, as even the best grism redshifts (from $e.g.$ \textit{Euclid}) have too low velocity resolution to identify true merger pairs (require $\lesssim$30\,km\,s$^{-1}$ resolution). \textbf{As such, there is no currently planned survey that will be able to robustly explore the mergers of galaxies below M* or beyond z$\sim$0.45.}

\vspace{1mm}

For a sample to make a transformative leap in understanding mergers at $z>0.3$, we define the following characteristics: \\

\vspace{-3mm}

1. Detect major merger (3:1) M* galaxies to $z$$\sim$1 (extending baseline of WAVES-deep by $\sim$4\,Gyr), \\

\vspace{-3mm}

2. Detect minor merger (10:1) M* galaxies to z$\sim$0.8 (extending baseline of WAVES-wide by $\sim$4\,Gyr), and \\

\vspace{-3mm}

 3. Probe a cosmologically representative volume at z$\sim$1 (100deg$^2$ covers $\sim$0.1\,Gpc$^3$ comoving volume in a $\Delta z$=0.1 bin at $z$=1, with $\sim$6\% cosmic variance error).  \\
 
 \vspace{-3mm}

Based on these requirements, we suggest a highly-complete redshift survey at 0.3$<$$z$$<$1.3 extending to i$\lesssim$24  and covering $\sim$100deg$^2$ (Fig 2. Right). In synergy with facilities such as LSST, \textit{Euclid}, Roman, SKA, this will enable a comprehensive understanding of stellar mass assembly and the astrophysical processes that regulate it. Currently there is no approved facility that could undertake such a survey in a reasonable timescale. Hence, we strongly argue in favour of the development of a 10m+ class, large field-of-view, highly-multiplexed multi-object spectroscopic facility. Finally, we also indicate that this study would be highly-beneficial in constraining simulations - which currently include many varied astrophysical models for both the impact of mergers of galaxies and feedback.

\vspace{1mm}


\vspace{-4mm}

 \begin{multicols}{3}   

\begin{scriptsize}

Casasola V.+, 2017, A\&A, 605, A18\\
Conselice C.~J., 2009, MNRAS, 394, 1956\\
Cook R.+, 2025, MNRAS, 539, 2829 \\
Darg D.~W.+, 2010, MNRAS, 401, 1552. \\
Davies L.+, 2015b, MNRAS, 452, 616 \\
Davies L.+, 2016a, MNRAS, 455, 4013 \\
Davies L.+, 2018, MNRAS, 480, 768 \\
Davies L.+, 2019b, MNRAS, 483, 5444 \\
Davies L.+, 2025b, MNRAS, 541, 573\\
Davies L.+, 2025c, MNRAS, 541, 3220 \\
Davies L.+, 2025e, MNRAS, 544, 3005 \\
Driver S.~P.+, 2011, MNRAS, 413, 971\\
Driver S.~P.+, 2019, Msngr, 175, 46 \\
Driver S.~P.+, 2022, MNRAS, 513, 439 \\
Ellison+ ,2019, MNRAS, 487, 2491 \\
Fuentealba-Fuentes+, 2025, MNRAS, 539, 1651\\
Greene J.+, 2022, arXiv, arXiv:2206.14908. \\
Hashemizadeh A.+, 2021, MNRAS, 505, 136 \\
Kennicutt R.~C., 1998, ApJ, 498, 541 \\
Lin L.+, 2008, ApJ, 681, 232 \\
Loubser+, 2022, MNRAS, 515, 1104\\ 
Maiolino R.+, 2020, Msngr, 180, 24\\
Martin+, 2018, MNRAS, 480, 2266 \\
Muzzin A.+, 2013, ApJ, 777, 18 \\
Neistein E., Dekel A., 2008, MNRAS, 388, 1792.\\
Pearson W.~J.+, 2019, A\&A, 631, A51\\
Popping G.+, 2014, MNRAS, 442, 2398 \\
Robotham A.+, 2014, MNRAS, 444, 3986\\
Sargent M.~T.+, 2007, ApJS, 172, 434\\
Sargent M.~T.+, 2025, OJAp, 8, 33\\
Scudder+, 2012, MNRAS, 426, 549\\
Stevens+, 2021, MNRAS, 502, 3158 \\
Springel V.+, 2005, Natur, 435, 629. \\
Thorne J.+, 2021, MNRAS, 505, 540 \\
Walter F.+, 2020, ApJ, 902, 111\\
Wilkinson S.+, 2022, MNRAS, 516, 4354\\

\end{scriptsize}

 \end{multicols}

\end{document}